\documentclass[twocolumn,superscriptaddress,letter]{revtex4-2}
\usepackage{amssymb}
\usepackage{graphicx}
\usepackage{amsmath}
\usepackage{amsthm}
\usepackage{multirow}
\usepackage{bm}
\usepackage[colorlinks=true] {hyperref}
\hypersetup{linkcolor=blue,filecolor=blue,urlcolor=blue,citecolor=green}
\usepackage{titlesec}
\usepackage{color}

\begin{document}
\title{Dissipation-induced Half Quantized Conductance in One-dimensional Topological Systems}

\author{Bozhen Zhou}
 \affiliation{Institute of Theoretical Physics, Chinese Academy of Sciences, Beijing 100190, China}

\author{Pan Zhang}
 \affiliation{Institute of Theoretical Physics, Chinese Academy of Sciences, Beijing 100190, China}

\author{Yucheng Wang}
\thanks{Corresponding author: wangyc3@sustech.edu.cn}
\affiliation{Shenzhen Institute for Quantum Science and Engineering,
	Southern University of Science and Technology, Shenzhen 518055, China}
\affiliation{International Quantum Academy, Shenzhen 518048, China}
\affiliation{Guangdong Provincial Key Laboratory of Quantum Science and Engineering, Southern University of Science and Technology, Shenzhen 518055, China}

\author{Chao Yang}
\thanks{Corresponding author: yangc@sustech.edu.cn}
\affiliation{Department of Physics, Southern University of Science and Technology, Shenzhen 518055, China}
\affiliation{Shenzhen Institute for Quantum Science and Engineering,
	Southern University of Science and Technology, Shenzhen 518055, China}

\begin{abstract}
Quantized conductance from topologically protected edge states is a hallmark of two-dimensional topological phases. In contrast, edge states in one-dimensional (1D) topological systems cannot transmit current across the insulating bulk, rendering their topological nature invisible in transport. In this work, we investigate the transport properties of the Su-Schrieffer-Heeger model with gain and loss, and show that the zero-energy conductance exhibits qualitatively distinct behaviors between the topologically trivial and nontrivial phases, depending on the hybridization and dissipation strengths. Crucially, we analytically demonstrate that the conductance can become half-quantized in the topologically nontrivial phase, a feature absent in the trivial phase. We further show that the half quantization predominantly originates from transport channels involving gain/loss and edge states. Our results uncover a new mechanism for realizing quantized transport in 1D topological systems and highlight the nontrivial role of dissipation in enabling topological signatures in open quantum systems.
\end{abstract}
\maketitle

Over the past decades, topological phases of matter have been one of the most actively studied fields in condensed matter physics and quantum simulation~\cite{RevModPhys.82.3045,RevModPhys.83.1057}. A key feature of topological phases is the presence of topologically protected edge states, whose transport properties are robust against disorder and defects. This gives rise to a variety of intriguing transport phenomena, 
such as the quantum Hall effect  and topological insulators~\cite{RevModPhys.82.3045,RevModPhys.83.1057,PhysRevLett.45.494,PhysRevLett.49.405,Kane2005PRL,Bernevig2006PRL,konig2007quantum}. Moreover, the protected boundary states directly result in quantized conductance~\cite{PhysRevLett.45.494,PhysRevLett.49.405}, whose extremely high precision allows for the accurate measurement of fundamental constants such as Planck's constant~\cite{Williams1998PRL}. However, topological features in one-dimensional (1D) systems remain elusive in transport, despite their well-established bulk classification and the presence of topologically protected boundary states. This is because the boundaries of a 1D chain are isolated zero-dimensional points that cannot form continuous current-carrying channels through the gapped bulk. As a result, the topological nature of 1D systems is difficult to detect via transport measurements.

While most studies of topological phases have focused on closed systems, dissipation is inevitable in realistic quantum platforms. Recent advances in engineering and controlling various forms of dissipation \cite{RevModPhys.85.553,barreiro2011open,muller2012engineered,kienzler2015quantum,dreon2022self,PhysRevLett.123.193605} have opened new opportunities to explore topological phenomena in open quantum systems.
While dissipation typically degrades quantum coherence and drives the system towards a structureless steady state \cite{PhysRevA.31.2403,RevModPhys.75.715}, recent research has revealed that dissipation can also be utilized to manipulate states of matter, leading to distinctive dynamics \cite{PhysRevLett.85.812,PhysRevX.9.031009,PhysRevLett.123.170401,PhysRevLett.122.050501,dogra2019dissipation, bouganne2020anomalous,PhysRevLett.130.200404}, unique steady states  \cite{PhysRevLett.125.240404,PhysRevLett.107.080503,PhysRevLett.101.105701,PhysRevLett.132.216301,syassen2008strong,PhysRevLett.102.144101, PhysRevLett.116.235302}, and specific transport properties \cite{PhysRevLett.132.136301,PhysRevLett.129.056802,PhysRevLett.105.190501,PhysRevLett.115.083601,PhysRevResearch.5.033095,Chao2024Arxiv,PhysRevLett.123.180402,PhysRevResearch.5.013195}. There has been growing interest in topological physics in open quantum systems, particularly in the topological characteristics of the Lindbladian~\cite{PhysRevLett.124.040401,PhysRevResearch.2.033428,PhysRevLett.127.250402,Lucas2023PRX}, non-equilibrium steady states~\cite{Tatsuhiko2020Science,Rakovszky2024PRX}, and the density matrix~\cite{Bardyn2018PRX,Wang2024SciPost}. Nevertheless, how dissipation affects the transport properties of topological systems remains poorly understood.

In this work, we explore the impact of on-site gain and loss on transport in a 1D topological system, using the Su-Schrieffer-Heeger (SSH) model~\cite{PhysRevLett.42.1698} as an example. We show that dissipation induces qualitatively distinct transport features between the topologically trivial and nontrivial phases. Specifically, the zero-energy conductance exhibits a half-quantized value that arises predominantly from transmission channels involving edge states and dissipative channels, rather than direct lead-to-lead tunneling. These findings reveal a new mechanism for realizing quantized transport in 1D topological systems.\\

\begin{figure}[t]
  \centering
  \includegraphics[width=1\linewidth]{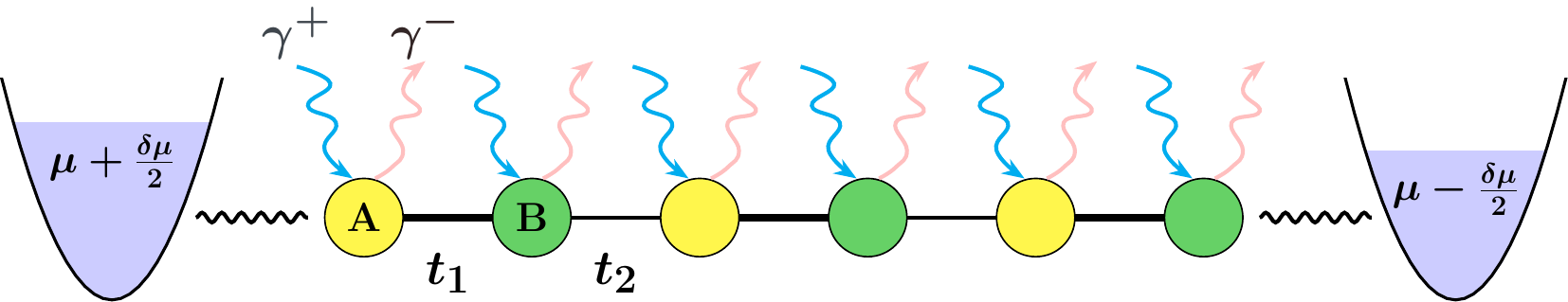}\\
  \caption{Schematic of the SSH model with on-site loss $\gamma^{-}$ and gain $\gamma^{+}$. The chain is coupled to two leads at its left and right ends, with chemical potentials $\mu+\frac{\delta\mu}{2}$ and $\mu-\frac{\delta\mu}{2}$, respectively.}\label{Fig1}
\end{figure}

\noindent{\large{\textbf{Model and Results}}}\\
\noindent \textbf{Dissipative SSH model}\\
\noindent We consider a two-terminal measurement setup for the dissipative SSH model, as shown in Fig. \ref{Fig1}, with dynamics described by the Lindblad master equation \cite{lindblad1976generators,breuer2002theory,How1993,Rivas_2014,Rivas_2010}:
\begin{equation}\label{1.1}
\frac{d\rho}{dt}=-i[\mathcal{H},\rho]+\sum_{\nu,j}(2\mathcal{L}_{\nu,j}\rho \mathcal{L}_{\nu,j}^{\dagger}-\{\mathcal{L}_{\nu,j}^{\dagger}\mathcal{L}_{\nu,j},\rho\}).
\end{equation}
Here, the total Hamiltonian is given by
\begin{equation}
\mathcal{H}=\mathcal{H}_{SSH}+\sum_{\alpha}(\mathcal{H}_{\alpha}+\mathcal{H}_{\alpha,SSH}),
\end{equation}
 where $\alpha=L,R$ denotes the left and right leads, respectively. The SSH model Hamiltonian~\cite{PhysRevLett.42.1698} is  $\mathcal{H}_{SSH}=\sum_{j}(t_1c_{j,A}^{\dagger}c_{j,B}+t_2c_{j,B}^{\dagger}c_{j+1,A}+h.c.)$, where $c_{j,A(B)}$ denotes the annihilation operator at sublattice $A$ ($B$) in unit cell $j$, and $t_1$ ($t_2$) is the intra-cell (inter-cell) hopping amplitude. The model undergoes a topological phase transition controlled by the ratio of $t_1$ and $t_2$: it is topologically nontrivial with degenerate zero-energy edge modes when $|t_1| < |t_2|$, and trivial when $|t_1| > |t_2|$ \cite{shen2012topological}. The leads are modeled as non-interacting fermionic baths described by  $\mathcal{H}_{\alpha}=\sum_{k}\epsilon_{\alpha,k}d_{\alpha,k}^{\dagger}d_{\alpha,k}$, where $d_{\alpha,k} (d^{\dagger}_{\alpha,k})$ is the annihilation (creation) operator for the $k$-th mode in the $\alpha$-th lead, and $\epsilon_{\alpha,k}$ represents the energy of this mode. The coupling between the leads and the SSH chain occurs at the two ends, and is described by $\mathcal{H}_{L,SSH}=\sum_{k}t_{L,1k}c_{1,A}^{\dagger}d_{L,k}+h.c.$ and $\mathcal{H}_{R,SSH}=\sum_{k}t_{R,Nk}c_{N,B}^{\dagger}d_{R,k}+h.c.$, where $t_{L,1k}$ and $t_{R,Nk}$ are the tunnel amplitudes, and $N$ is the number of unit cells, so the system size is $2N$.
The dissipative processes are modeled by Lindblad operators: $\mathcal{L}_{1,j}=\sqrt{\gamma_{j}^{-}}c_{j}$ for on-site loss, and $\mathcal{L}_{2,j}=\sqrt{\gamma_{j}^{+}}c_{j}^{\dagger}$ for on-site gain, where  $\gamma_{j}^{-}$ and $\gamma_{j}^{+}$ denote the local loss and gain strengths at site $j$, respectively. Throughout this work, we focus on the zero-temperature regime, where the lead chemical potentials are set to $\mu_L=\mu+\frac{\delta\mu}{2}$ and $\mu_R=\mu-\frac{\delta\mu}{2}$.\\

\noindent \textbf{Steady-state current and differential
conductance}\\
\noindent In the presence of gain and loss, the steady-state current flowing out of the left lead ($J_{L}$) and into the right lead ($J_{R}$) may differ. The net current is defined as $J=\frac{1}{2}(J_L+J_R)$. Using the Lindblad-Keldysh formalism \cite{sieberer2016keldysh}, $J$ can be written as \cite{Chao2024Arxiv} (See Supplemental Material)
\begin{align}\label{1.2}
	J=\int_{-\infty}^{\mu}\frac{d\omega}{h}\tau_{0,l}(\omega) -\int_{\mu}^{\infty}\frac{d\omega}{h}\tau_{0,g}(\omega) +\int_{\mu-\frac{\delta\mu}{2}}^{\mu+\frac{\delta\mu}{2}}\frac{d\omega}{h}\tau_1(\omega).
\end{align}
The first two terms, which are independent of the bias $\delta\mu$, arise solely from dissipation and represent the contributions from loss and gain processes. These are characterized by the transmission functions
$\tau_{0,l}=\mathrm{Tr}[(\bm{\Gamma}_{L}-\bm{\Gamma}_R)\mathbf{G}^{\mathcal{R}}\mathbf{P}\mathbf{G}^{\mathcal{A}}]$ and $\tau_{0,g}=\mathrm{Tr}[(\bm{\Gamma}_{L}-\bm{\Gamma}_R)\mathbf{G}^{\mathcal{R}}\mathbf{Q}\mathbf{G}^{\mathcal{A}}]$. Here, bold symbols denote matrices: $\mathbf{P}$ and $\mathbf{Q}$ are diagonal matrices encoding loss and gain, with elements $P_{jk}=\gamma_j^- \delta_{jk}$ and $Q_{jk}=\gamma_j^+ \delta_{jk}$. The retarded Green's function is given by $\mathbf{G}^{\mathcal{R}}=(\mathbf{G}^{\mathcal{A}})^{\dagger} =(\omega-\mathbf{H}_{SSH}+i\mathbf{P}+i\mathbf{Q}-\bm{\Sigma}_{L}^{\mathcal{R}}-\bm{\Sigma}_{R}^{\mathcal{R}})^{-1}$, with $\bm{\Gamma}_{\alpha}= i(\bm{\Sigma}_{\alpha}^{\mathcal{R}}-\bm{\Sigma}_{\alpha}^{\mathcal{A}})$ the hybridization matrices. In the wide-band limit, the self-energies simplify to $(\bm{\Sigma}_{L}^{\mathcal{R}})_{1A,1A}=(\bm{\Sigma}_{R}^{\mathcal{R}})_{NB,NB}=-\frac{i\Gamma}{2}$, where $\Gamma$ is the hybridization strength to the leads.
The third term in Eq.~(\ref{1.2}), referred to as the response current $\delta J_1$, determines the differential conductance. This current depends explicitly on the chemical potential bias $\delta\mu$ and is mediated by the gain–loss matrix $\mathbf{P} + \mathbf{Q}$ through the transmission function $\tau_1=\frac{1}{2}\mathrm{Tr}[\bm{\Gamma}_{L}\mathbf{G}^{\mathcal{R}}\bm{\Gamma}_{R}\mathbf{G}^{\mathcal{A}}+ \bm{\Gamma}_{R}\mathbf{G}^{\mathcal{R}}\bm{\Gamma}_{L}\mathbf{G}^{\mathcal{A}}]+\theta(\omega-\mu)\mathrm{Tr}[ \bm{\Gamma}_{L}\mathbf{G}^{\mathcal{R}}(\mathbf{P}+\mathbf{Q})\mathbf{G}^{\mathcal{A}}]+\theta(\mu-\omega)\mathrm{Tr}[ \bm{\Gamma}_{R}\mathbf{G}^{\mathcal{R}}(\mathbf{P}+\mathbf{Q})\mathbf{G}^{\mathcal{A}}]$\cite{Chao2024Arxiv}.  To measure the differential conductance, which depends only on the third term in Eq.~(\ref{1.2}), one must subtract the contributions of the first two terms from the measured current. These contributions can be obtained by measuring the current at $\delta\mu = 0$, as the total current under this condition consists solely of the first two terms in Eq.~(\ref{1.2}).
When the system and the applied gain/loss are reflection symmetric, the first two terms vanish naturally \cite{Chao2024Arxiv}. The SSH Hamiltonian is known to preserve inversion symmetry. Under uniform gain and loss, i.e., $\gamma_j^- \equiv \gamma^-$ and $\gamma_j^+ \equiv \gamma^+$, the dissipation matrices $\mathbf{P}$ and $\mathbf{Q}$ also respect inversion symmetry.
These symmetry relations cause the first two terms in Eq.~(\ref{1.2}) to vanish \cite{Chao2024Arxiv}, leaving the measured current determined solely by the third term.

\begin{figure}[t]
	\centering
	\includegraphics[width=1\linewidth]{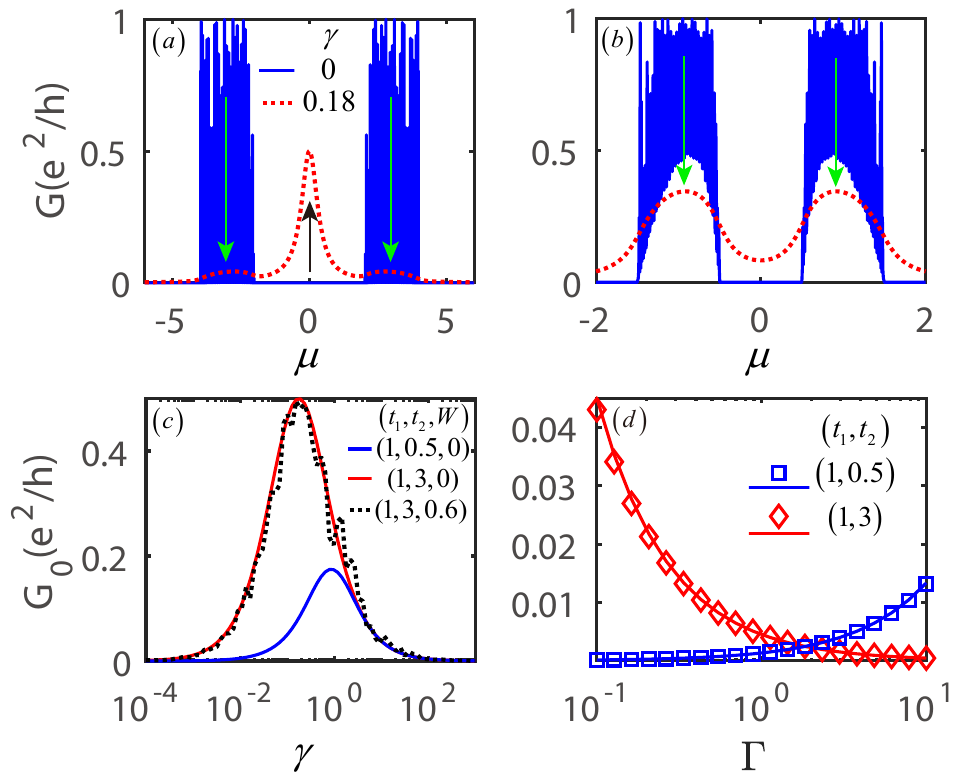}\\
	\caption{The differential conductance as a function of various parameters. (a)(b) Differential conductance (in units of $e^2/h$) as a function of chemical potential $\mu$ for $N=40$ and $\Gamma=0.4$: (a) Topological nontrivial phase ($t_1=1$, $t_2=3$); (b) trivial phase ($t_1=1$, $t_2=0.5$). (c) Zero energy conductance versus dissipation strength $\gamma$ at fixed $\Gamma=0.4$. (d) Zero energy conductance versus hybridization strength  $\Gamma$ at fixed $\gamma=0.001$. Diamonds and squares are numerical results; solid curves show the analytical approximation from Eq. (\ref{2.2}).}\label{Fig2}
\end{figure}

In the linear response regime ($\delta\mu\rightarrow 0$), the differential conductance is defined as~\cite{Chao2024Arxiv} (See Supplemental Material) 
\begin{align}
	& G = e^2 \frac{\delta J_1}{\delta\mu} = \frac{e^2}{h} \mathrm{Tr}[\mathbf{g}(\mu)]= \frac{e^2}{h} (\mathrm{Tr}[\mathbf{g}_1] + \mathrm{Tr}[\mathbf{g}_2]), \label{1.3} \\
	&\mathrm{Tr}[\mathbf{g}_1] = \frac{1}{2} \mathrm{Tr}[\bm{\Gamma}_L \mathbf{G}^{\mathcal{R}} \bm{\Gamma}_R \mathbf{G}^{\mathcal{A}} + \bm{\Gamma}_R \mathbf{G}^{\mathcal{R}} \bm{\Gamma}_L \mathbf{G}^{\mathcal{A}}], \nonumber \\
	&\mathrm{Tr}[\mathbf{g}_2] = \frac{1}{2} \mathrm{Tr}[(\bm{\Gamma}_L + \bm{\Gamma}_R) \mathbf{G}^{\mathcal{R}} (\mathbf{P} + \mathbf{Q}) \mathbf{G}^{\mathcal{A}}], \nonumber
\end{align}
where $\mathrm{Tr}[\mathbf{g}(\mu)]$ represents the total normalized conductance at energy $\mu$, which consists of two contributions: the first term, $\mathrm{Tr}[\mathbf{g}_1]$, corresponds to the standard Landauer transmission between the two leads; the second term, $\mathrm{Tr}[\mathbf{g}_2]$, accounts for indirect transmission pathways involving gain and loss, i.e., processes in which particles are scattered into or out of the leads through the dissipative reservoirs.
In the absence of gain and loss, $\mathrm{Tr}[\mathbf{g}_2] = 0$, and the conductance is fully determined by the Landauer term. In this case, transport is ballistic within the energy band, and the conductance peaks at the quantum limit $G = e^2/h$, as shown in both the topologically nontrivial [Fig. \ref{Fig2}(a)] and trivial [Fig. \ref{Fig2}(b)] phases.
However, once dissipation is introduced (with total strength $\gamma = \gamma^+ + \gamma^-$), particles can scatter into the local reservoirs, resulting in a finite lifetime and suppressed transmission. Consequently, the conductance peaks are reduced, as indicated by the green arrows in Fig. \ref{Fig2}(a) and (b). Moreover, dissipation broadens the spectral features outside the energy band, rendering $G(\mu)$ a smooth function of $\mu$.

\noindent \textbf{Zero-energy conductance}\\  
\noindent A particularly interesting feature appears at zero energy ($\mu = 0$) in the topologically nontrivial phase, where a sharp conductance peak emerges, as marked by the black arrow in Fig. \ref{Fig2}(a). 
To gain insight into the transport characteristics at zero energy, we analyze how the zero-energy conductance, denoted as $G_0\equiv G(\mu=0)$, depends on the hybridization strength $\Gamma$ and the dissipation strength $\gamma$ in different topological phases. Using the transfer matrix method, $G_0$ can be expressed as (See Supplemental Material)
\begin{equation}\label{2.1}
G_0=\frac{e^2}{h}\frac{2\gamma\Gamma x}{(\frac{\Gamma}{2t_2}x+2\gamma t_2)^2}+O(\lambda^{-N}),
\end{equation}
where $x=\sqrt{[\gamma^2+(t_1+t_2)^2][\gamma^2+(t_1-t_2)^2]}-\gamma^2-t_1^2+t_2^2$, and $\lambda>1$ ensures the second term vanishes exponentially with system size. In the weak dissipation limit ($\gamma\ll|t_1-t_2|,\Gamma$), $G_0$ simplifies to
\begin{equation}\label{2.2}
\begin{split}
& G_0\approx\frac{e^2}{h}\frac{4\gamma t_2^2}{\Gamma(t_2^2-t_1^2)},~~\text{topological~nontrivial}, \\
& G_0\approx\frac{e^2}{h}\frac{\gamma\Gamma}{t_1^2-t_2^2},~~~~~~\text{topological~trivial}.
\end{split}
\end{equation}
In both the topologically nontrivial and trivial phases,  $G_0$ increases linearly with $\gamma$ [Fig.~\ref{Fig2}(c)]. However, its dependence on $\Gamma$ exhibits an opposite trend: $G_0$ decreases with increasing $\Gamma$ in the nontrivial phase, while it increases in the trivial phase 
[Fig.~\ref{Fig2}(d)], providing a clear signature that  distinguishes the two phases.
In the strong dissipation limit ($\gamma \to \infty$), $G_0$ decays as $\Gamma/\gamma$, becoming independent of $t_1$ and $t_2$, as shown in Fig.~\ref{Fig2}(c). This decay behavior reflects the quantum Zeno effect \cite{misra1977zeno,PhysRevLett.122.040402}.

At intermediate dissipation strength, the conductance reaches a maximum at a critical value $\gamma_c$ [Fig. \ref{Fig2}(c)]. In the topological nontrivial phase, the maximum conductance can reach the theoretical upper bound of half the conductance quantum \cite{note1}: $G_0(\gamma_c)=e^2/2h$,
and the corresponding critical dissipation strength is given by
\begin{equation}\label{2.4}
\gamma_c=\frac{\Gamma^2-4t_2^2+\sqrt{(\Gamma^2-4t_2^2)^2+16\Gamma^2(t_2^2-t_1^2)}}{4\Gamma}.
\end{equation}
From Eq. (\ref{2.4}), it is evident that a positive solution for $\gamma_c$ always exists in the topologically nontrivial phase. In contrast, in the trivial phase, the conductance remains strictly below $e^2/2h$, as illustrated by the blue curve in Fig. \ref{Fig2}(c) \cite{note2}. We note that the conductance peak at $e^2/2h$ persists even under random fluctuations in the gain and loss strengths, as shown by the dashed lines in Fig.~\ref{Fig2}(c). Specifically, disorder is introduced via $\gamma_j = \gamma_0(1 + W_j)$, where $W_j$ is uniformly distributed in the interval $[-W, W]$, and the parameter $W\in [0,1]$ quantifies the disorder strength. This result demonstrates the robustness of the half-quantized conductance against imperfections in the gain/loss configuration.

\begin{figure}[t]
  \centering
  \includegraphics[width=1\linewidth]{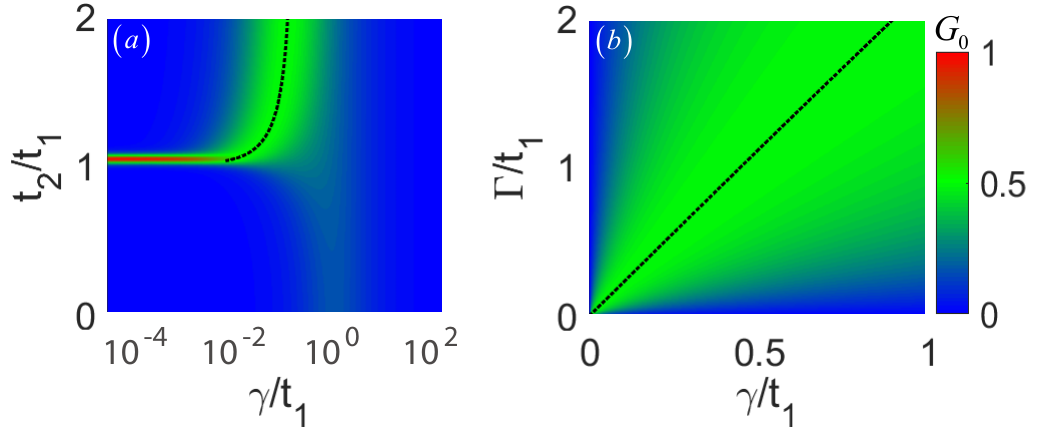}\\
  \caption{The emergence of half-quantized zero energy conductance. (a) and (b) show $G_0$ versus $\gamma/t_1$ and $t_2/t_1$, and versus $\gamma/t_1$ and $\Gamma/t_1$, respectively, with $N=40$, $\Gamma/t_1=0.4$ in (a), and $t_2/t_1=3$ in (b). The black dashed curve marks the critical dissipation strength $\gamma_c$ derived from Eq. (\ref{2.4}).}\label{Fig3}
\end{figure}

To provide an intuitive understanding, we present how the half-quantized conductance emerges and evolves as functions of the dissipation strength $\gamma$, the hopping ratio $t_2/t_1$, and the hybridization strength $\Gamma$. Fig. \ref{Fig3}(a) shows the zero-energy conductance $G_0$ as a function of $\gamma/t_1$ and $t_2/t_1$. In the weak dissipation limit ($\gamma \to 0$), the conductance approaches $e^2/h$ when $t_1 = t_2$, as the spectrum becomes gapless at $\mu = 0$, allowing nearly ballistic transport. For $t_1 \ne t_2$, the conductance is significantly reduced due to the presence of an energy gap. The black dashed curve indicates the critical dissipation strength $\gamma_c$ in the topologically nontrivial regime. Near the phase boundary ($t_2\approx t_1$), $\gamma_c$ exhibits dramatic variations with the hopping strength. In contrast, far from the phase boundary ($t_2\gg t_1$), $\gamma_c \sim \Gamma/2$ becomes approximately linear in $\Gamma$ and independent of the hopping amplitudes. Fig. \ref{Fig3}(b) plots $G_0$ as a function of $\gamma$ and $\Gamma$. The conductance peak of $e^2/2h$ can be realized by tuning either $\gamma$ or $\Gamma$. As $\Gamma$ increases, the conductance remains near its peak value across a broader range of $\gamma$. Therefore, the half-quantized conductance can be more easily observed experimentally by either fixing a large hybridization $\Gamma$ and tuning the gain/loss strength $\gamma$, or by fixing a large $\gamma$ and adjusting $\Gamma$.\\

\begin{figure}[t]
	\centering
	\includegraphics[width=1\linewidth]{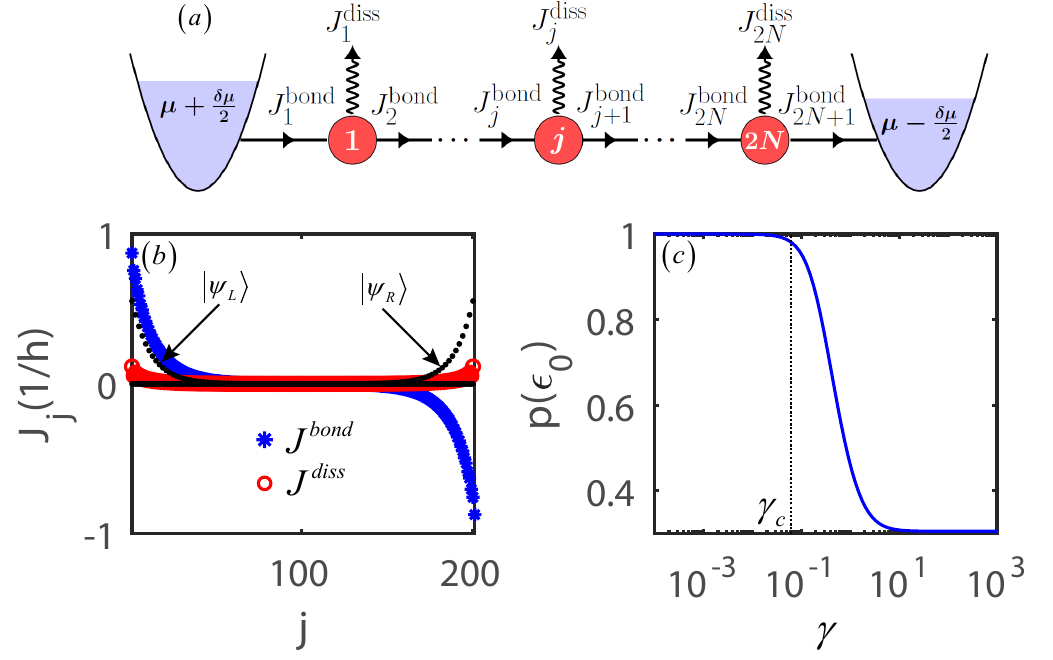}\\
	\caption{The contribution to the half-quantized zero energy conductance carried by topological edge states. (a) Schematic of the bond current $J_{j}^{\mathrm{bond}}$ and the dissipative current $J_{j}^{\mathrm{diss}}$.
		(b) Spatial distribution of $J_{j}^{\mathrm{bond}}$ and $J_{j}^{\mathrm{diss}}$ (in units of $1/h$) in the topologically nontrivial phase with $t_1 = 1$,  $t_2 = 1.2$, and $N = 100$. Black dotted curves show the edge-state wavefunctions $|\psi_L\rangle$ and $|\psi_R\rangle$. Parameters: $\gamma^{-} = 0.062$, $\gamma^{+} = 0$, $\mu = 0$, $\delta\mu = 0.01$, and $\Gamma = 0.4$.
		(c) The ratio function $p(\epsilon_0)$ as a function of $\gamma$. The black dotted line marks the critical dissipation strength $\gamma_c = 0.062$ as given in Eq.~(\ref{2.4}). }\label{Fig4}
\end{figure}

\noindent \textbf{Bond current and energy-resolved conductance}\\
\noindent To understand how the topological phase influences the emergence of half-quantized conductance, we now analyse the spatial distribution of current at the critical dissipation strength $\gamma_c$. To this end, we examine the dynamics of the local particle number operator $\mathcal{N}_j=c_j^{\dagger}c_j$, which evolves according to the Lindblad equation in the Heisenberg picture (See Supplemental Material):
\begin{equation}\label{2.5}
\frac{d\langle\mathcal{N}_j\rangle}{dt}=\frac{i}{\hbar}\langle[\mathcal{H},c_j^{\dagger}c_j]\rangle +\sum_{\nu}\langle 2\mathcal{L}_{\nu,j}^{\dagger}\mathcal{N}_j \mathcal{L}_{\nu,j}-\{\mathcal{L}_{\nu,j}^{\dagger}\mathcal{L}_{\nu,j},\mathcal{N}_j\}\rangle.
\end{equation}
The first term represents coherent particle transport along the chain, while the second term accounts for incoherent exchange with the local loss ($\nu=1$) and gain ($\nu=2$) channels. The bond current $J_{j}^{\mathrm{bond}}$ is defined through the continuity relation: $\frac{i}{\hbar}\langle[\mathcal{H},c_j^{\dagger}c_j]\rangle=J_{j}^{\mathrm{bond}}-J_{j+1}^{\mathrm{bond}}$. 
At the system boundaries, the bond currents satisfy $J_{1}^{\mathrm{bond}} = J_L$ and $J_{2N+1}^{\mathrm{bond}} = J_R$. The dissipative current $J_{j}^{\mathrm{diss}}$, defined as the contribution from the Lindblad operators (i.e., the second term in Eq.~(\ref{2.5})), quantifies the net particle flow into the environment due to local gain or loss processes. In the steady state, where the local particle number remains constant, i.e., $\frac{d\langle \mathcal{N}_j \rangle}{dt} = 0$, the continuity equation implies a simple relation: $J_{j}^{\mathrm{diss}} = J_{j}^{\mathrm{bond}} - J_{j+1}^{\mathrm{bond}}$, as illustrated in Fig.~\ref{Fig4}(a). Fig.~\ref{Fig4}(b) shows the spatial distribution of $J_{j}^{\mathrm{bond}}$ and $J_{j}^{\mathrm{diss}}$ at the critical dissipation $\gamma_c$. It is evident that the bond current is concentrated near the boundaries, where it flows from the leads into the chain, while remaining small in the bulk. Similarly, the dissipative current $J_{j}^{\mathrm{diss}}$ is also significant only near the boundaries. Notably, both currents are spatially confined within the localization length of the edge states, as shown by the black dotted wavefunction envelopes in Fig.~\ref{Fig4}(b). This indicates that the conductance is primarily determined by the edge states. 

To quantify the contribution of edge states, we define the energy-resolved conductance at zero energy as $G_0(\epsilon_n)=\frac{e^2}{h} \langle\psi_n|\mathbf{g}(\omega=0)|\psi_n\rangle$, where $\epsilon_n$ and $|\psi_n\rangle$ are the eigenvalues and eigenstates of the SSH Hamiltonian, satisfying $\mathcal{H}_{SSH} |\psi_n\rangle = \epsilon_n |\psi_n\rangle$. The total conductance is the sum over all eigenstate contributions, $G_0 = \sum_n G_0(\epsilon_n)$. We define the ratio function
\begin{equation}\label{2.7}
	p(\epsilon_n) = \frac{G_0(\epsilon_n)}{G_0},
\end{equation}
which quantifies the fractional contribution of state $|\psi_n\rangle$ to the total conductance, with $0 \leq p(\epsilon_n) \leq 1$. The contribution from the edge state is denoted as $p(\epsilon_0)$, where $\epsilon_0 = 0$.
In Fig.~\ref{Fig4}(c), we plot $p(\epsilon_0)$ as a function of the dissipation strength $\gamma$. For $\gamma < \gamma_c$, we find $p(\epsilon_0) \approx 1$, indicating that the conductance is almost entirely dominated by the edge state. As $\gamma$ increases beyond $\gamma_c$, a sharp drop in $p(\epsilon_0)$ is observed, revealing that bulk states begin to make significant contributions to the conductance.

The origin of the half quantization can be understood by analyzing the individual transmission processes. Since there is only one conducting channel connecting the left and right leads, both the left-to-right and right-to-left transmission functions ($\mathrm{Tr}[\bm{\Gamma}_L \mathbf{G}^{\mathcal{R}} \bm{\Gamma}_R \mathbf{G}^{\mathcal{A}}]$ and $\mathrm{Tr}[\bm{\Gamma}_R \mathbf{G}^{\mathcal{R}} \bm{\Gamma}_L \mathbf{G}^{\mathcal{A}}]$) remain below one. Similarly, transmissions from either lead to the reservoirs via gain or loss ($\mathrm{Tr}[\bm{\Gamma}_{L(R)} \mathbf{G}^{\mathcal{R}} (\mathbf{P} + \mathbf{Q}) \mathbf{G}^{\mathcal{A}}]$) are individually bounded by one-half due to their single-sided nature.
It follows from Eq.~(\ref{1.3}) that $\mathrm{Tr}[\mathbf{g}_1] \leq 1$ and $\mathrm{Tr}[\mathbf{g}_2] \leq 1/2$. At zero energy, the bulk states are gapped, and the Landauer term decays exponentially with system size. Thus, in the thermodynamic limit, the conductance is dominated by $\mathrm{Tr}[\mathbf{g}_2]$, and the total conductance saturates at $G \leq e^2/2h$. In the topologically nontrivial phase, the presence of edge states enables efficient tunneling between the leads and the gain/loss channels, supporting a substantial conductance despite the presence of a bulk gap.\\

\noindent{\large{\textbf{Discussion}}}\\
\noindent We have investigated how on-site gain and loss affect the transport properties of the SSH model. Our results show that the conductance is suppressed within the band but enhanced outside, with a pronounced zero-energy peak appearing in the topologically nontrivial phase. We derive the expression for the zero-energy conductance $G_0$, revealing that in the weak dissipation regime, $G_0$ increases with the hybridization $\Gamma$ in the trivial phase but decreases in the nontrivial phase, offering a characteristic feature for experimentally distinguishing whether a system is topologically trivial or nontrivial. Moreover, we identify a half-quantized conductance peak in the nontrivial phase as $\gamma$ varies, which is another distinct transport feature that differentiates it from the trivial phase. We analytically derive the relationship among $\gamma$, $\Gamma$, $t_1$, and $t_2$ that must be satisfied for $G_0$ to be half-quantized. Furthermore, we show that the half-quantized value remains robust against disordered dissipation strength and predominantly originates from topological edge states. Our work demonstrates that gain/loss can induce quantized transport in 1D topological systems, analogous to the quantum Hall effect in closed two-dimensional systems. These findings open new perspectives for exploring transport in dissipative topological systems and suggest potential applications in designing precise quantum transport devices and conducting accurate measurements in 1D systems. Future work could explore the impact of dissipation on the transport properties of higher-dimensional topological systems.\\

\noindent {\large{\textbf{References}}}\\
\bibliography{refs1}

\begin{thebibliography}{59}
\bibitem [{not()}]{note1}%
  \BibitemOpen
  \href@noop {} {}\bibinfo {note} {For a spin-degenerate system, the
  conductance is inherently associated with a multiplicity of two.}\BibitemShut
  {Stop}%
\bibitem [{In the topological trivial phase, the hybridization should satisfy
  $\Gamma<2|t_1|+2\sqrt{t_1^2-t_2^2}$ to ensure no bound states exist at the
  boundary()}]{note2}%
  \BibitemOpen
  In the topological trivial phase, the hybridization should satisfy
  $\Gamma<2|t_1|+2\sqrt{t_1^2-t_2^2}$ to ensure no bound states exist at the
  boundary,\ \href@noop {} {}\BibitemShut {NoStop}%
\end{thebibliography}
 
\noindent {\large{\textbf{Acknowledgments}}}\\
This work is supported by the National Key R\&D Program of China under Grant No.2022YFA1405800, the Key-Area Research and Development Program of Guangdong Province (Grant No.2020B0303010001, Grant No.2019ZT08X324, No.2019CX01X042, No.2019B121203002). B.Z. acknowledges support from Project 12447153 of the National Natural Science Foundation of China.

\noindent {\large{\textbf{Author contributions}}}\\
B.Z. and C.Y. conceived the idea for this work and performed the study. Y.W. and C.Y. wrote the major part of the paper. All authors contributed to the preparation of the manuscript.\\

\noindent {\large{\textbf{Competing interests}}}\\
The authors declare no competing interests.\\

\end{document}